\documentclass[sn-nature]{sn-jnl}% Style for submissions to Nature Portfolio journals
\usepackage{graphicx}%
\usepackage{multirow}%
\usepackage{amsmath,amssymb,amsfonts}%
\usepackage{amsthm}%
\usepackage{mathrsfs}%
\usepackage[title]{appendix}%
\usepackage{xcolor}%
\usepackage{textcomp}%
\usepackage{manyfoot}%
\usepackage{booktabs}%
\usepackage{algorithm}%
\usepackage{algorithmicx}%
\usepackage{algpseudocode}%
\usepackage{listings}%
\usepackage{url}
\usepackage{hyperref}
\usepackage[encapsulated]{CJK}
\usepackage[numbers]{natbib}
\usepackage{ragged2e}
\usepackage{setspace}
\usepackage{multibib}
\newcites{supp}{Methods References}
\theoremstyle{thmstyleone}%
\theoremstyle{thmstyletwo}%

\theoremstyle{thmstylethree}%
\def\aj{\rmfamily{Astron. J.~}}           % Astronomical Journal
\def\apj{\rmfamily{Astrophys. J.~}}         % Astrophysical Journal
\def\apjl{\rmfamily{Astrophys. J. L.~}}        % Astrophysical Journal, Letters
\def\apjs{\rmfamily{Astrophys. J. S.~}}       % Astrophysical Journal, Supplement

\def\aap{\rmfamily{Astron. Astrophys.~}}        % Astronomy and Astrophysics

\def\araa{\rmfamily{ARA\&A~}}     % Annual Review of Astronomy and Astrophysics
\def\mnras{\rmfamily{Mon. Not. R. Astron. Soc.~}}     % Monthly Notices of the Royal Astron. Soc.
\def\pasp{\rmfamily{PASP~}}       % Publications of the Astron. Soc. o/t Pacific
\def\nat{\rmfamily{Nature~}}      % Nature
\begin{document}

\title{A high black hole to host mass ratio in a lensed AGN in the early Universe}

%%=============================================================%%
%% Prefix	-> \pfx{Dr}
%% GivenName	-> \fnm{Joergen W.}
%% Particle	-> \spfx{van der} -> surname prefix
%% FamilyName	-> \sur{Ploeg}
%% Suffix	-> \sfx{IV}
%% NatureName	-> \tanm{Poet Laureate} -> Title after name
%% Degrees	-> \dgr{MSc, PhD}
%% \author*[1,2]{\pfx{Dr} \fnm{Joergen W.} \spfx{van der} \sur{Ploeg} \sfx{IV} \tanm{Poet Laureate} 
%%                 \dgr{MSc, PhD}}\email{iauthor@gmail.com}
%%=============================================================%%

\author*[1]{\fnm{Lukas J.} \sur{Furtak}}\email{furtak@post.bgu.ac.il}
\author[2]{\fnm{Ivo} \sur{Labb\'{e}}}
\author[1]{\fnm{Adi} \sur{Zitrin}}
\author[3]{\fnm{Jenny E.} \sur{Greene}}
\author[4]{\fnm{Pratika} \sur{Dayal}}
\author[5]{\fnm{Iryna} \sur{Chemerynska}}
\author[4]{\fnm{Vasily} \sur{Kokorev}}
\author[6,7]{\fnm{Tim B.} \sur{Miller}}
\author[3]{\fnm{Andy D.} \sur{Goulding}}
\author[8]{\fnm{Anna} \sur{de Graaff}}
\author[9]{\fnm{Rachel} \sur{Bezanson}}
\author[10]{\fnm{Gabriel B.} \sur{Brammer}}
\author[11]{\fnm{Sam E.} \sur{Cutler}}
\author[12,13,14]{\fnm{Joel} \sur{Leja}}
\author[15]{\fnm{Richard} \sur{Pan}}
\author[9]{\fnm{Sedona H.} \sur{Price}}
\author[12,13,14]{\fnm{Bingjie} \sur{Wang}}
\author[11]{\fnm{John R.} \sur{Weaver}}
\author[11,10]{\fnm{Katherine E.} \sur{Whitaker}}
\author[5]{\fnm{Hakim} \sur{Atek}}
\author[16]{\fnm{\'Akos} \sur{Bogd\'an}}
\author[5]{\fnm{St\'{e}phane} \sur{Charlot}}
\author[17]{\fnm{Emma} \sur{Curtis-Lake}}
\author[6]{\fnm{Pieter} \sur{van Dokkum}}
\author[18]{\fnm{Ryan} \sur{Endsley}}
\author[19]{\fnm{Robert} \sur{Feldmann}}
\author[20,21]{\fnm{Yoshinobu} \sur{Fudamoto}}
\author[18]{\fnm{Seiji} \sur{Fujimoto}}
\author[2]{\fnm{Karl} \sur{Glazebrook}}
\author[22]{\fnm{St\'{e}phanie} \sur{Juneau}}
\author[15]{\fnm{Danilo} \sur{Marchesini}}
\author[23]{\fnm{Micheal V.} \sur{Maseda}}
\author[24]{\fnm{Erica} \sur{Nelson}}
\author[25,10]{\fnm{Pascal A.} \sur{Oesch}}
\author[26]{\fnm{Ad\`{e}le} \sur{Plat}}
\author[9]{\fnm{David J.} \sur{Setton}}
\author[26]{\fnm{Daniel P.} \sur{Stark}}
\author[22,26]{\fnm{Christina C.} \sur{Williams}}

\affil[1]{Physics Department, Ben-Gurion University of the Negev, P.O. Box 653, Beer-Sheva 8410501, Israel}
\affil[2]{Centre for Astrophysics and Supercomputing, Swinburne University of Technology, Melbourne, VIC 3122, Australia}
\affil[3]{Department of Astrophysical Sciences, Princeton University, 4 Ivy Lane, Princeton, NJ 08544, USA}
\affil[4]{Kapteyn Astronomical Institute, University of Groningen, P.O. Box 800, 9700 AV Groningen, The Netherlands}
\affil[5]{Institut d’Astrophysique de Paris, CNRS, Sorbonne Universit\'{e}, 98bis Boulevard Arago, 75014, Paris, France}
\affil[6]{Department of Astronomy, Yale University, New Haven, CT 06511, USA}
\affil[7]{Center for Interdisciplinary Exploration and Research in Astrophysics (CIERA) and Department of Physics and Astronomy, Northwestern University, 1800 Sherman Ave, Evanston IL 60201, USA}
\affil[8]{Max-Planck-Institut f\"{u}r Astronomie, K\"{o}nigstuhl 17, 69117 Heidelberg, Germany}
\affil[9]{Department of Physics and Astronomy and PITT PACC, University of Pittsburgh, Pittsburgh, PA 15260, USA}
\affil[10]{Cosmic Dawn Center (DAWN), Niels Bohr Institute, University of Copenhagen, Jagtvej 128, K{\o}benhavn N, DK-2200, Denmark}
\affil[11]{Department of Astronomy, University of Massachusetts, Amherst, MA 01003, USA}
\affil[12]{Department of Astronomy \& Astrophysics, The Pennsylvania State University, University Park, PA 16802, USA}
\affil[13]{Institute for Computational \& Data Sciences, The Pennsylvania State University, University Park, PA 16802, USA}
\affil[14]{Institute for Gravitation and the Cosmos, The Pennsylvania State University, University Park, PA 16802, USA}
\affil[15]{Department of Physics and Astronomy, Tufts University, 574 Boston Ave., Medford, MA 02155, USA}
\affil[16]{Center for Astrophysics $\vert$ Harvard \& Smithsonian, 60 Garden Street, Cambridge, MA 02138, USA}
\affil[17]{Centre for Astrophysics Research, Department of Physics, Astronomy and Mathematics, University of Hertfordshire, Hatfield AL10
9AB, UK}
\affil[18]{Department of Astronomy, The University of Texas at Austin, Austin, TX 78712, USA}
\affil[19]{Institute for Computational Science, University of Zurich, CH-8057 Zurich, Switzerland}
\affil[20]{Waseda Research Institute for Science and Engineering, Faculty of Science and Engineering, Waseda University, 3-4-1 Okubo, Shinjuku, Tokyo 169-8555, Japan}
\affil[21]{National Astronomical Observatory of Japan, 2-21-1, Osawa, Mitaka, Tokyo, Japan}
\affil[22]{NSF’s National Optical-Infrared Astronomy Research Laboratory, 950 N. Cherry Avenue, Tucson, AZ 85719, USA}
\affil[23]{Department of Astronomy, University of Wisconsin-Madison, 475 N. Charter St., Madison, WI 53706 USA}
\affil[24]{Department for Astrophysical and Planetary Science, University of Colorado, Boulder, CO 80309, USA}
\affil[25]{Department of Astronomy, University of Geneva, Chemin Pegasi 51, 1290 Versoix, Switzerland}
\affil[26]{Steward Observatory, University of Arizona, 933 N Cherry Ave, Tucson, AZ 85721 USA}

\maketitle

\textbf{
Early JWST observations have uncovered a new population of red sources that might represent a previously overlooked phase of supermassive black hole growth \cite{kocevski23,matthee23,labbe23}. One of the most intriguing examples is an extremely red, point-like object that was found to be triply-imaged by the strong lensing (SL) cluster Abell~2744 \cite{furtak23d}. Here we present deep JWST/NIRSpec observations of this object, Abell2744-QSO1. The spectroscopy confirms that the three images are of the same object, and that it is a highly reddened ($A_V\simeq3$) broad emission line Active Galactic Nucleus (AGN) at a redshift of $z_{\mathrm{spec}}=7.0451\pm0.0005$. From the width of H$\beta$ ($\mathrm{FWHM}=2800\pm250\,\frac{\mathrm{km}}{\mathrm{s}}$) we derive a black hole mass of $M_{\mathrm{BH}}=4_{-1}^{+2}\times10^7\,\mathrm{M}_{\odot}$. We infer a very high ratio of black hole to galaxy mass of at least 3\,\%, an order of magnitude more than is seen in local galaxies \cite{bennert11}, and possibly as high as 100\,\%. The lack of strong metal lines in the spectrum together with the high bolometric luminosity ($L_{\mathrm{bol}}=(1.1\pm0.3)\times10^{45}\,\frac{\mathrm{erg}}{\mathrm{s}}$) indicate that we are seeing the black hole in a phase of rapid growth, accreting at 30\,\% of the Eddington limit. The rapid growth and high black hole to galaxy mass ratio of A2744-QSO1 suggest that it may represent the missing link between black hole seeds \cite{volonteri21} and the first luminous quasars \cite{fan22}.
}

\bigskip

% Introduction
A2744-QSO1 \cite{furtak23d} was discovered in JWST/\textit{Near Infrared Camera} \cite{rieke23} (NIRCam) pre-imaging from the \textit{Ultra-deep NIRSpec and NIRCam ObserVations before the Epoch of Reionization} \cite{bezanson22} (UNCOVER) program, behind the \textit{Hubble Frontier Fields} \cite{lotz17} (HFF) SL cluster Abell~2744 \cite{abell89} ($z=0.308$; A2744 hereafter), known to be capable of significantly magnifying distant background objects \cite{furtak23c}. It had been identified as a UV-faint ($M_{\mathrm{UV},1450}=-16.98\pm0.09$), lensed high-redshift candidate \cite{atek14a}, but remained an otherwise inconspicuous source in previous $0.5-1.5\,\mu$m observations with the \textit{Hubble Space Telescope} (HST). The NIRCam $1-5\,\mu$m imaging revealed it to be an unusually red (e.g. F277W $-$ F444W$=2.63\pm0.10$), unresolved point-source triply imaged by the foreground cluster \cite{furtak23d}, with the high-redshift nature of the source further supported geometrically by the lensing configuration. Its unique HST and JWST spectral energy distribution (SED), combined with the point-like light-profile, hinted that the source may be a highly reddened AGN at redshift $z_{\mathrm{phot}}\simeq7-8$ \cite{furtak23d,labbe23}.

The three images of A2744-QSO1 were recently targeted with the \textit{Near Infrared Spectrograph} \cite{jakobsen22} (NIRSpec) aboard JWST as part of the UNCOVER program (see Fig.~1). Stacking the spectra of the three images of A2744-QSO1, we obtain one $\sim38$\,h spectrum, the deepest spectrum so far of any single high-redshift object observed with JWST. After correcting for gravitational magnifications \cite{furtak23c} of $\mu_{\mathrm{A}}=6.2_{-0.2}^{+0.5}$, $\mu_{\mathrm{B}}=7.3_{-1.5}^{+0.1}$ and $\mu_{\mathrm{C}}=3.5_{-0.2}^{+0.2}$, this corresponds to a 1.5\,micron continuum depth of $\sim31.0$\,AB at $5\sigma$, equivalent to integrating for $\gtrsim1700$\,h in a blank field.

% Evidence for AGN
The spectrum shown in Fig.~2 features several strong Hydrogen emission lines -- a prominent Balmer-series with H$\alpha$, H$\beta$, H$\gamma$, and H$\delta$, as well as Lyman-$\alpha$ -- along with several faint metal lines. We give a full list of detected emission lines in Tab.~\ref{tab:emission lines}. The broad Hydrogen emission lines securely confirm this object at $z_{\mathrm{spec}}=7.0451\pm0.0005$ and provide unambiguous evidence that A2744-QSO1 is an AGN. The broad-line width is measured by fitting the H$\beta$ line, which is the clearest and most isolated since H$\alpha$ is situated on the edge of the detector. We thus find an H$\beta$ full width at half-maximum (FWHM) of $\mathrm{FWHM}=2800\pm250\,\frac{\mathrm{km}}{\mathrm{s}}$, well-resolved, even at the relatively low resolution of the prism (see Fig.~2). Such line widths are not seen in star-forming galaxies but routinely observed in the broad-line regions of AGN, where gas clouds orbit the supermassive black hole \cite{hao05b}. An additional fit to the [O\,\textsc{iii}]\,$\lambda5008$\AA\ line clearly shows that it is narrow and un-resolved as opposed to the broad Balmer lines. Assuming an SMC extinction law \cite{gordon03}, the Balmer line ratio (H$\alpha$ to H$\beta$) indicates a strong dust attenuation of $A_V=3.0\pm0.5$. Note that flatter attenuation curves yield higher $A_V$. The \textit{Atacama Large Millimeter/sub-millimeter Array} (ALMA) 1.2\,mm observations of A2744 \cite{fujimoto23} firmly rule out dusty star-formation powering the Balmer lines. If caused by star-formation, the emission lines would imply an on-going star-formation rate (SFR) of $\psi\sim40\,\frac{\mathrm{M}_{\odot}}{\mathrm{yr}}$ and a corresponding 1.2\,mm flux of $\sim1$\,mJy, whereas no emission is observed to $<0.1$\,mJy (at $3\sigma$). We can therefore clearly rule out the possibility of our source being a star-forming galaxy. However, we cannot entirely rule out that the observed rest-frame UV emission is at least partly of stellar origin. In that case, assuming the entire UV emission comes from stars, we can estimate the stellar mass by assuming a constant star-formation history (SFH) and a formation redshift of $z=12$ and obtain $M_{\star}\sim10^{8.2}\,\mathrm{M}_{\odot}$.

% BH mass + luminosity
From the width of the H$\beta$ line (see Fig.~2), we derive the black hole mass using standard scaling relations \cite{greene05}, and obtain $M_{\mathrm{BH}}=4_{-1}^{+2}\times10^7\,\mathrm{M}_{\odot}$. We further derive a bolometric luminosity of $L_{\mathrm{bol}}=(1.1\pm0.3)\times10^{45}\,\frac{\mathrm{erg}}{\mathrm{s}}$ (see Fig.~3) based on the emission line luminosity. Both $M_{\mathrm{BH}}$ and $L_{\mathrm{bol}}$ are corrected for magnification. The black hole must accrete at a rate of $\dot{M}\sim0.1\,\frac{\mathrm{M}_{\odot}}{\mathrm{yr}}$ to maintain this luminosity, which roughly corresponds to 30\,\% of the Eddington limit -- set by the balance between gravity and radiation pressure assuming spherical accretion ($L_{\mathrm{bol}}/L_{\mathrm{Edd}}\sim0.3$).

We can also place a limit on the hidden stellar mass of the host galaxy based on the light profile. All three images are point-sources, despite lensing magnifications differing by a factor 2, which suggests that the intrinsic size of the source is smaller than the point spread function (PSF) of the highest-magnification image. To derive formal upper limits on the size, we fit a S\'{e}rsic \cite{sersic63} model to the source in the (highest-resolution) F115W-band, finding the source to be unresolved with a size $r_e<30$\,pc (95\,\% upper limit) after correcting for the lensing shear (equivalent to less than half the size of the PSF). Adopting an upper stellar density limit equal to that of the densest star clusters \cite{vanzella23} or the densest elliptical galaxy progenitors \cite{baggen23} known to date, $\Sigma_{\star}\sim5\times10^5\,\frac{\mathrm{M}_{\odot}}{\mathrm{pc}^2}$, we can derive an upper limit for how much stellar mass may be contained within $r_e<30$\,pc, thus obtaining $M_{\star}<1.4\times10^{9}\,\mathrm{M}_{\odot}$. Our prior photometric fit including ALMA 1.1\,mm constraints returns an intrinsic stellar mass upper limit that is consistent with this estimate \cite{furtak23d,labbe23}. The implied $M_{\mathrm{BH}}-M_{\star}$-ratio of $>3\,\%$ is extreme compared to local values ($M_{\mathrm{BH}}/M_{\star}\sim0.1\,\%$) but similar to some other recently discovered high-redshift black holes \cite{izumi21}.

% On the specialness of our source
The spectrum of A2744-QSO1 (Fig.~2) exhibits some distinct characteristics compared to typical AGN \cite{vandenberk01} and other moderate-luminosity $z>4$ AGN discovered with JWST \cite{harikane23}. The optical continuum is very red, and the metal or high-ionization lines (e.g. the C\,\textsc{iv}\,$\lambda\lambda1548,1550$\AA, C\,\textsc{iii}]\,$\lambda\lambda1907,1909$\AA, Mg\,\textsc{ii}\,$\lambda\lambda2796,2803$\AA\ and the [O\,\textsc{iii}]$\lambda\lambda4959,5007$\AA-doublets) are very weak compared to the Hydrogen lines (see Fig.~2 and Tab.~\ref{tab:emission lines}). Furthermore, using the latest deep X-ray imaging obtained by the \textit{Chandra X-ray Observatory}, we find that all three images are undetected in the X-rays. The combined $3\sigma$ upper limit on the rest-frame 40\,keV X-ray luminosity of our source is $L_{\rm X,40\,keV}<3\times10^{43}\,\mathrm{erg}\,\mathrm{s}^{-1}$, at least $10\times$ weaker than expected based on the UV/optical. This intrinsic X-ray faintness of A2744-QSO1 might suggest a powerful dusty wind, as are seen in low-mass rapidly accreting black holes at low redshift \cite{veilleux16}. Such dusty wind-dominated sources are rare at cosmic noon \cite{banerji15}, however, and we do not detect any direct evidence for a strong wind in the current data. On the other hand, many black holes radiating at or above their Eddington limits display these unique properties \cite{leighly07}, perhaps because of changes in the structure of their accretion disk \cite{abramowicz88}. Our estimated Eddington ratio of 0.3 would also be consistent with these characteristics.

% Number density
A2744-QSO1 represents one of the first examples of a heavily reddened broad-line AGN (``red dots'') at high redshift \cite{furtak23d}, although hints existed also in prior work \cite{fujimoto22}. These objects are too UV-faint to have been discovered by past ground-based photometric searches \cite{matsuoka23}, but many similar compact red objects have been detected in JWST surveys since \cite{harikane23,matthee23,labbe23}, as well as other high-redshift AGN \cite{kocevski23,larson23,harikane23,maiolino23a,maiolino23b}, including behind A2744 \cite{goulding23}. In fact, the UNCOVER ``red dot'' sample \cite{labbe23} contains a second object at $z_{\mathrm{spec}}=7.04$ \cite{greene23}, consistent with A2744-QSO1, suggesting these objects may be clustered. Indeed, three out of seven similar objects found in a blank field study \cite{matthee23} were also at similar redshifts to each other. Nevertheless, A2744-QSO1 remains a unique case: As the highest-redshift quasar-like object multiply imaged by a galaxy cluster known to date, it affords exquisite signal-to-noise for such a low-luminosity object.

We derive the spatial number density of A2744-QSO1-type objects by taking the sample of ``red dots'' detected in UNCOVER \cite{labbe23,greene23} and verifying how many of them are confirmed at the same redshift and UV luminosity with the UNCOVER spectroscopy and updated SL model. This yields a number density of $\phi=(7.3\pm1.7)\times10^{-5}\,\mathrm{Mpc}^{-3}\,\mathrm{mag}^{-1}$, which is consistent with what we found previously based on the UNCOVER photometry \cite{labbe23} ($\phi=2.5_{-2.1}^{+2.6}\times10^{-4}\,\mathrm{Mpc}^{-3}\,\mathrm{mag}^{-1}$) and also with what has been found for reddened $z>5$ AGN in JWST blank fields \cite{matthee23}. While this represents only a few percent of galaxies at this epoch \cite{atek18}, it is $\sim100$ times more numerous than the faintest UV-selected AGN \cite{matsuoka23}. Because the UV-luminosity of A2744-QSO1 is only $M_{\mathrm{UV},1450}=-16.98\pm0.09$, finding this extremely red source at high redshifts required the unprecedented infrared-sensitivity of JWST combined with gravitational lensing. As can be seen in the upper panels of Fig.~3, while A2744-QSO1 is very faint in rest-frame UV compared to other high-redshift AGN, it is nevertheless typical in terms of bolometric luminosity.

% Simulations and theory
A2744-QSO1 is a significant outlier, both in terms of its $M_{\mathrm{BH}}-M_{\star}$-ratio (which is a lower limit as we do not detect the host galaxy) and the expected number density when compared to theoretical models such as the semi-analytic \texttt{DELPHI} model \cite{dayal19,piana21,habouzit22}. Assuming black hole seeds with a seed mass of $150\,\mathrm{M}_\odot$ originating from metal-free stars, the \texttt{DELPHI} model predicts that black holes as massive as A2744-QSO1 should be hosted by galaxies with stellar masses of about $M_{\star}\simeq10^{10.5}\,\mathrm{M}_\odot$, i.e. much lower $M_{\mathrm{BH}}-M_{\star}$ ratios (see Fig.~3), and it under-predicts the number density of A2744-QSO1-like objects. Hydrodynamic simulations make similar predictions at $z\sim6$ \cite{habouzit22}, although typically starting with heavier seeds ($10^{4-6}\,\mathrm{M}_{\odot}$). In general, the models fail to predict the high number density associated with A2744-QSO1, which can only be reproduced by the most extreme models without feedback (Fig.~3). This is in conflict with the hints of strong outflows in our source discussed above.

% Wrap-up and conclusion
The discovery of A2744-QSO1 shows that our understanding of black hole growth in the early Universe is still incomplete. The photometric samples suggest high number densities, such that a few percent of galaxies at $z>7$ harbor reddened and rapidly growing black holes. Deep spectroscopic follow-up with JWST will be needed to confirm these number densities and elucidate these objects' properties. The hints of extreme $M_{\mathrm{BH}}-M_{\star}$-ratios, of a few percent or more, will put strong new constraints on when black holes form relative to their host galaxies. Black holes that outgrow their galaxies put strain on AGN evolution models and might hint at significant dark matter accretion onto primordial seeds \cite{delaurentis22} or at hyper-compact star-bursting clusters \cite{kroupa20} in the early Universe. There are many outstanding puzzles about these objects remaining that JWST is only beginning to address. For A2744-QSO1 in particular, higher-resolution spectroscopy in the rest-frame UV could reveal the possible presence of a dusty outflow in this source, while observations at longer wavelengths could help to better constrain the mass of its host galaxy.

\begin{table}[h!]
\centering
\caption{\justifying \textbf{Emission lines detected in the combined spectrum of A2744-QSO1.} The stacked spectrum is presented in Fig.~2. Columns represent, from left to right: Emission line designation, rest-frame wavelength, observed wavelength, measured line flux (not corrected for magnification), rest-frame equivalent width (if measured). Upper limits are at $3\sigma$. For Lyman-$\alpha$, the continuum is based on the extrapolated rest-frame UV continuum assuming a power-law.}
\begin{tabular}{lcccc}
Emission Line            &   $\lambda_0$ [\AA]    &   $\lambda_{\mathrm{obs}}$ [$\mu$m]    &   $\mu F_{\mathrm{line}}~\left[10^{-20}\,\frac{\mathrm{erg}}{\mathrm{s}\,\mathrm{cm}^2}\right]$  &  $\mathrm{EW}_0$~[\AA]\\\hline
Ly$\alpha$                      &   $1216$          &        $0.978$           &   $236.8\pm18.8$       & $65.5\pm12.1$ \\
Mg\,\textsc{ii}                 &   $2796$, $2803$  &        $2.252$           &   $<11$                & -\\
He\,\textsc{i}                  &   $3889$          &        $3.129$           &   $18.3\pm3.3$         & -\\   
{[Ne\,\textsc{iii}]}            &   $3968$          &        $3.192$           &   $15.2\pm2.5$         & -\\
H$\delta$                       &   $4102$          &        $3.300$           &   $19.0\pm2.4$         & -\\
H$\gamma^{\mathrm{a}}$          &   $4341$          &        $3.492$           &   $34.5\pm3.4$         & -\\
H$\beta$                        &   $4863$          &        $3.912$           &   $100.8\pm3.3$        & $42.5\pm2.1$ \\
{[O\,\textsc{iii}]}             &   $4960$          &        $3.990$           &   $5.0\pm2.6$          & -\\
{[O\,\textsc{iii}]}             &   $5008$          &        $4.029$           &   $15.5\pm2.6$         & -\\
Si\,\textsc{ii}$^{\mathrm{b}}$  &   $6347$          &       $5.106$            &   $21.3\pm3.2$         & -\\
{[N\,\textsc{ii}]}              &   $6550$          &       $5.270$            &   $76.9\pm19.1$        & -\\
H$\alpha$                       &   $6565$          &       $5.282$            &   $753.2\pm29.1$       & $158.0\pm11$ \\\hline
\end{tabular}
\par\smallskip
\begin{flushleft}
    $^{\mathrm{a}}$\, Blended with [O\,\textsc{iii}]$\lambda4363$\AA.
    \par $^{\mathrm{b}}$\, Blended with [O\,\textsc{i}]$\lambda6364$\AA\ and [Fe\,\textsc{x}]$\lambda6375$\AA.
\end{flushleft}
\label{tab:emission lines}
\end{table}

\begin{figure}[h!]
    \centering
    \includegraphics[width=\textwidth]{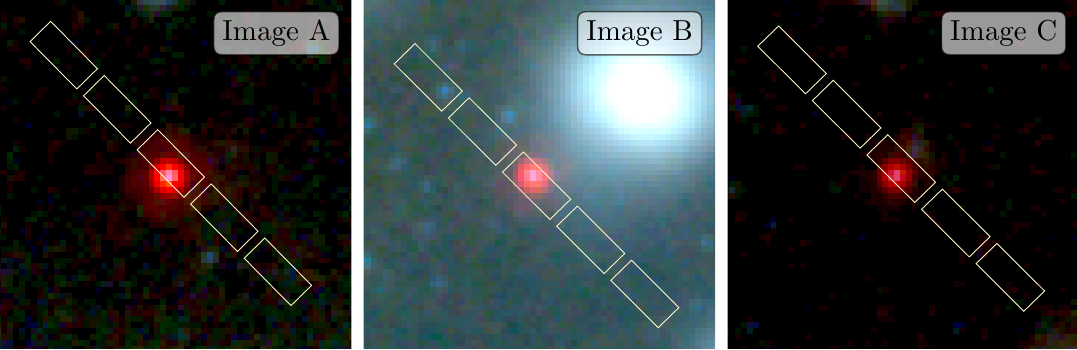}
    \caption{\textbf{Composite-color image cutouts of the three images of A2744-QSO1.} The $2.4''\times2.4''$ cutouts of the UNCOVER JWST-NIRCam composite-color image \cite{furtak23c} (Blue: F115W+F150W, Green: F200W+F277W, Red: F2356W+F410M+F444W) show the three images of A2744-QSO1 overlaid with the NIRSpec MSA slitlets of one of our MSA masks as an example (the three images were targeted with several masks and the exact position of the object in each slit may slightly shift).}
    \label{fig:cutouts}
\end{figure}

\begin{figure}[h!]
    \centering
    \includegraphics[width=\textwidth]{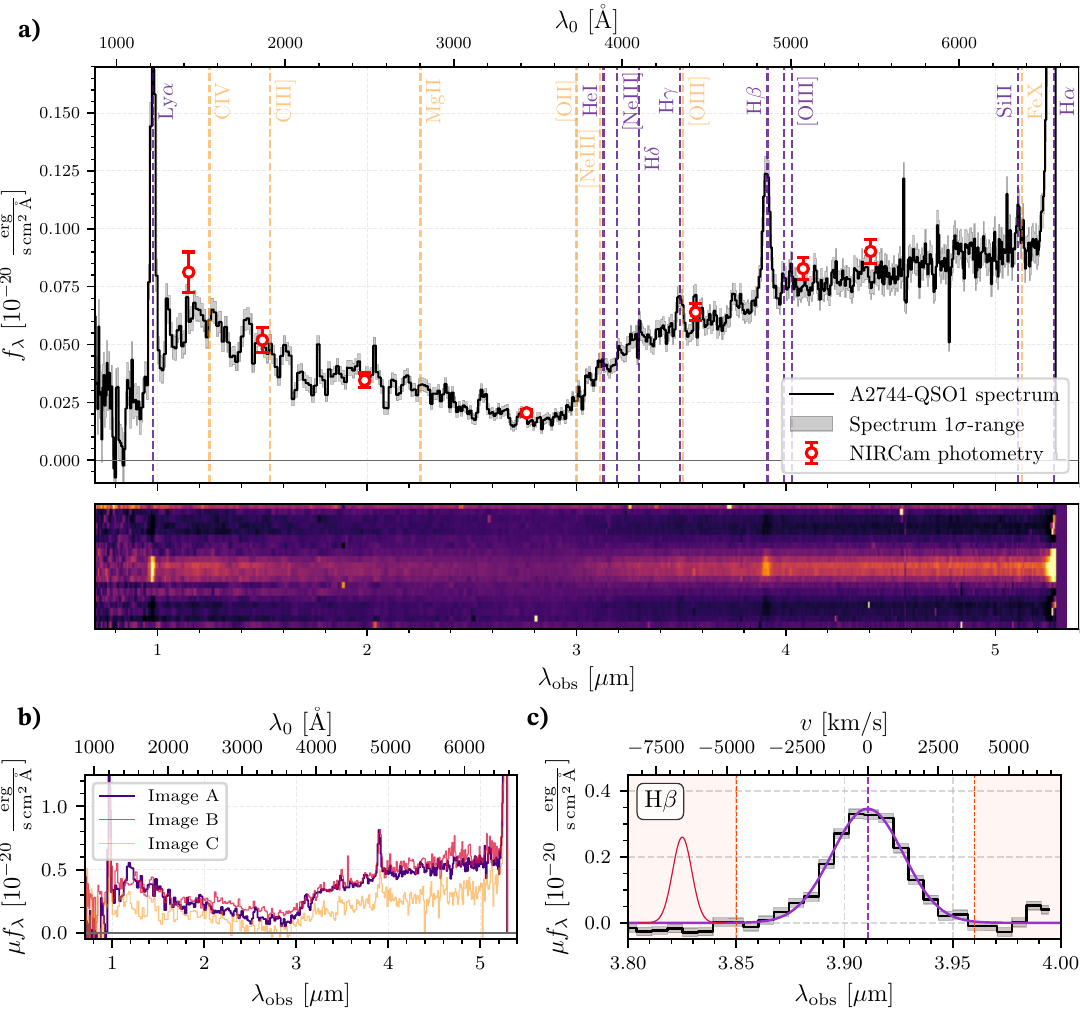}
    \caption{\textbf{NIRSpec-prism spectrum of A2744-QSO1.} Panel a) shows the de-magnified stacked spectrum (black), equivalent to $\sim1700$\,h integration time on target once the lensing is factored-in, and its $1\sigma$-uncertainty range (grey). In addition to its red rest-frame optical continuum \cite{furtak23d}, the source clearly displays strong Hydrogen emission lines in the Balmer-series and Lyman-$\alpha$. The emission lines allow us to precisely measure the spectroscopic redshift at $z_{\mathrm{spec}}=7.0451\pm0.0005$. We overlay the NIRCam photometry measured previously \cite{furtak23d} as red circles. Securely identified and detected ($\geq3\sigma$) emission lines are marked in purple and selected non-detected lines are shown in orange. Panel b) shows the individual (magnified) spectra of the three images. All three images perfectly align in wavelength space, thus confirming the triply-imaged nature of our object. Panel c) shows the continuum-subtracted H$\beta$ line (black) and our Gaussian fit to the line (purple). The red shaded areas delimit the region in which the fit is performed and the dark red curve illustrates the NIRSpec-prism LSF at the wavelength of H$\beta$. All errors shown represent $1\sigma$ uncertainties.}
    \label{fig:spectrum}
\end{figure}

\begin{figure}[h!]
    \centering
    \includegraphics[width=\textwidth]{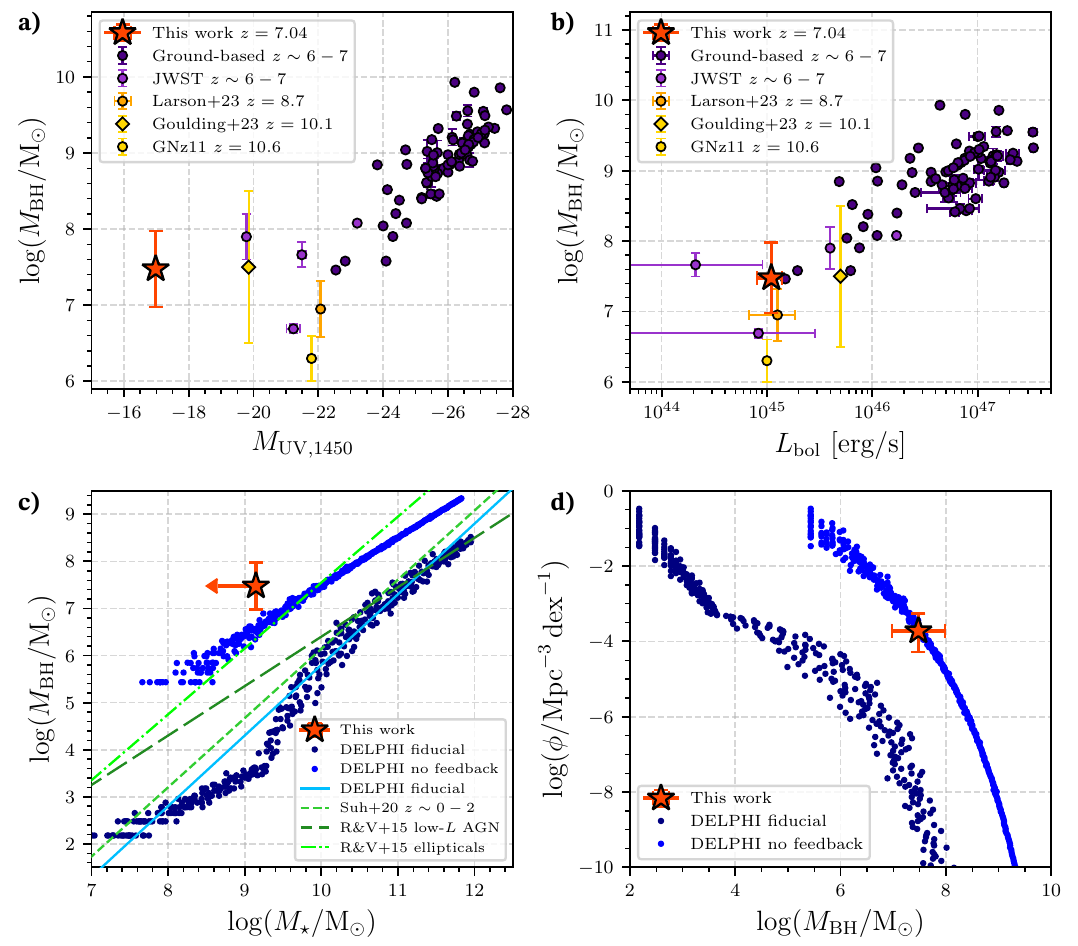}
    \caption{\textbf{A2744-QSO1 compared to other broad-line AGN.} The top row shows A2744-QSO1 (red star) in the black hole mass-luminosity space of high-redshift broad-line AGN. Purple circles show UV-bright AGN observed from the ground \cite{yang21,izumi21}, the violet circles show AGN detected with JWST at $z\sim6-7$ \cite{fujimoto22,harikane23,maiolino23b}, and the orange and yellow symbols show recently confirmed AGN at higher redshifts \cite{larson23,goulding23,maiolino23a}. While A2744-QSO1 is notably fainter in the UV than other AGN recently detected with JWST (a) because of its strong dust attenuation, it is typical in the $M_{\mathrm{BH}}-L_{\mathrm{bol}}$ plane (b). The bottom row shows A2744-QSO1 in $M_{\mathrm{BH}}-M_{\star}$-space (c) and in number-densities (d) compared to the \texttt{DELPHI} simulations \cite{dayal19,piana21} (blue) and AGN observed at low redshift \cite{reines15,suh20} (green). The dark blue points show the default \texttt{DELPHI} runs, which predict much higher host stellar masses and much lower number density for a black hole of this mass. The light blue points show the models with feedback turned off, in which case the models can approach the observations of A2744-QSO1. Error bars represent $1\sigma$ uncertainties.}
    \label{fig:Mbh_vs_L}
\end{figure}

\clearpage

% ------------------ Methods --------------------------------------------
\section*{Methods}

Throughout this work, we assume a standard flat $\Lambda$CDM cosmology with $H_0=70\,\frac{\mathrm{km}}{\mathrm{s}\,\mathrm{Mpc}}$, $\Omega_{\Lambda}=0.7$, and $\Omega_\mathrm{m}=0.3$. Magnitudes are quoted in the AB system \cite{oke83} and uncertainties represent $1\sigma$ ranges unless stated otherwise.

\subsection*{JWST, ALMA, and \textit{Chandra} observations}
The UNCOVER observations used in this work comprise both ultra-deep JWST \cite{gardner23,mcelwain23} imaging and spectroscopy of the A2744-field with NIRCam \cite{rieke23} and NIRSpec \cite{jakobsen22,boeker23} to $5\sigma$-depths of $29.9$\,magnitudes each \cite{bezanson22,weaver23}. Since gravitational magnification on average adds $\sim2$\,magnitudes to the depth in the SL regime, UNCOVER reaches effective depths of $\sim32$\,AB once the magnification is factored-in. The NIRCam data cover a total area of $\sim45\,\mathrm{arcmin}^2$ in six broad- and one medium-band filter: F115W, F150W, F200W, F277W, F356W, F410M and F444W. The data were reduced and drizzled into mosaics with the \texttt{Grism redshift and line analysis software for space-based slitless spectroscopy} \cite{grizli23} (\texttt{grizli}; \texttt{v1.6.0}). The UNCOVER mosaics, which also include other public JWST imaging data (i.e. the JWST-GLASS imaging \cite{treu22} and DDT program 2756; PI: W.~Chen; as well as archival HST imaging) and were made publicly available on the UNCOVER website. The photometry for all three images of A2744-QSO1 can be found in Table~1 of the discovery paper for A2744-QSO1 \cite{furtak23d}.

\noindent The NIRSpec component of the UNCOVER survey consists of seven micro-shutter array (MSA) \cite{ferruit22} pointings on the A2744-field with various exposure times between $\sim3$\,h and $\sim17$\,h, totalling 24\,h NIRSpec time and 680 targeted objects. The observations follow a 2-point dither pattern and a 3-shutter slit-let nod pattern at a position angle of $266^{\circ}$. The observations were taken from July~31 - August~2, 2023, using the low-resolution prism disperser, which covers the wavelength range $0.7-5.3\,\mu\mathrm{m}$ at spectral resolutions $R\sim30-300$. The exposure times for the three images of A2744-QSO1 ranged from $9.3-16.4$\,h. The prism data were reduced with \texttt{MSAEXP} \cite{msaexp22} (\texttt{v0.6.10}). The code works on individual frames to correct for $1/f$-noise, find and mask snowballs, and remove the bias. The pipeline first applies a WCS, and then identifies each slit and applies the flat-fielding and photometric corrections. The 2D-slits are then extracted and drizzled onto a common grid. The details of the observational setup and reduced data products, planned for public release before JWST Cycle~3, will be presented in Price et al. (in prep.). All three images of A2744-QSO1 were included in the MSA designs with cumulated exposure times $17.4$\,h, $9.9$\,h and $14.7$\,h for images A, B and C respectively.

\noindent In addition, A2744 is part of the \textit{ALMA Frontier Fields} survey \cite{gonzales-lopez17}, which obtained first deep 1.1\,mm imaging of the cluster core. These data have now recently been complemented with additional observations covering the entire UNCOVER footprint (Program~ID: 2022.1.00073.S; PI: S.~Fujimoto), achieving 1$\sigma$ depths down to 32.7\,$\mu$Jy/beam over an area of $\sim$24\,arcmin$^{2}$. The UNCOVER ALMA data were also publicly released and described in detail in Fujimoto et al. \cite{fujimoto23}.

\noindent A2744 has also been the subject of sensitive X-ray follow-up observations using the \textit{Chandra X-ray observatory}. The 1.25\,Ms of \textit{Advanced CCD Imaging Spectrometer} (ACIS) data used here were previously reduced, analyzed and presented in the supplementary materials section of Bogd\'{a}n et al. \cite{bogdan23}. We use these data products to perform forced photometry at the locations of the three images of A2744-QSO1. Images A and B are heavily contaminated by the foreground X-ray gas associated with A2744 itself. Image C provides the cleanest (lowest-background) region to perform an X-ray analysis of A2744-QSO1. However, we find no significant detection of an X-ray point source. We place a $3\sigma$ upper limit of $<12.8$ photons in the $2-10$\,keV energy band, i.e. a flux limit of $f_{\rm X,2-10\,keV}<2.8\times10^{-16}\,\frac{\mathrm{erg}}{\mathrm{s}\,\mathrm{cm}^2}$ assuming a typical AGN X-ray power-law slope of $\Gamma=1.9$. This is equivalent to a de-magnified X-ray luminosity of $L_{\rm X,40\,keV}<5\times10^{43}\,\frac{\mathrm{erg}}{\mathrm{s}}$ at a rest-frame energy of $\sim40$\,keV given the high redshift of A2744-QSO1. Stacking and de-magnifying all three images produces a marginal increase in sensitivity, owing to the increased magnification but higher backgrounds of images A and B of $L_{\rm X,40\,keV}<3\times10^{43}\,\frac{\mathrm{erg}}{\mathrm{s}}$.

\subsection*{Gravitational lensing}
Throughout this work we use the most recent SL model of A2744 by Furtak et al. \cite{furtak23c}, which is based on UNCOVER data and also publicly available on the UNCOVER website. The model is slightly updated here with one new multiple image system and an additional spectroscopic redshift \cite{bergamini23b}. The model was constructed using a new implementation of the Zitrin-parametric code \cite{zitrin15a,pascale22,furtak23c} , which is not coupled to a fixed grid resolution and thus capable of high resolution results. For the model, the cluster galaxies are modeled each as a double Pseudo Isothermal Ellipsoid \cite{eliasdottir07} (dPIE), scaled according to its luminosity using common scaling relations motivated by the fundamental plane. Three of the brightest cluster galaxies (BCGs), as well as several other galaxies situated close to multiple images, are modeled independently. We use 421 cluster galaxies throughout the UNCOVER $\sim45\,\mathrm{arcmin}^2$ field-of-view, more than half of which are spectroscopically confirmed \cite{bergamini23a}. The rest were chosen based on the cluster's red-sequence, identified in a color-magnitude diagram based on the HST/ACS F606W- and F814W-bands bracketing their 4000\,\AA-break. Five cluster-scale dark matter (DM) halos are employed, each modeled as a Pseudo Isothermal Elliptical Mass Distribution \cite{jaffe83,keeton01a} (PIEMD). These DM halos are centered on the five BCGs but for three of them, the central position can be iterated for in the minimization process within a couple of arc-seconds around their respective BCG's position. As constraints, 141 multiple images (belonging to 48 sources) are used, 96 of which have spectroscopic redshifts. The vast majority of these is situated in the main cluster core just like A2744-QSO1. The minimization is performed in the source plane using a $\sim10^5$-step MCMC process. For the minimization and uncertainty estimation, we use a positional uncertainty of $0.5"$ for each multiple image. The resulting model has a very good image reproduction RMS of $\Delta_{\mathrm{RMS}}=0.51"$ in the lens plane. We infer magnifications and 95\% confidence intervals of $\mu_{\mathrm{A}}=6.15~[5.76,6.92]$, $\mu_{\mathrm{B}}=7.29~[5.11,7.65]$ and $\mu_{\mathrm{C}}=3.55~[3.31,3.80]$ for images A, B and C respectively. These are only very slightly different from the magnifications presented in Furtak et al. \cite{furtak23d} and are in generally good agreement with the photometric flux ratios.

\subsection*{Spectroscopic analysis}
We use the image multiplicity of A2744-QSO1 to achieve maximum signal-to-noise by stacking the weighted, local-background subtracted and co-added spectra of the three images. The spectrum is then extracted with \texttt{MSAEXP} using an optimal extraction \cite{horne86} and collapsed to a 1D-spectrum. Taking the magnification into account, the effective exposure time of this stacked spectrum is of about 1700\,h. The flux-calibration is then performed by bootstrapping to the total photometry of image A, which has the cleanest photometric measurement \cite{furtak23d}. Wavelength-dependent slit-losses are modeled with a first-order polynomial for each MSA by convolving with the filter band-pass and solving for the polynomial coefficients by comparing with the photometry. While our stacking and flux-calibration procedures are optimal for maximizing the signal-to-noise and measuring properties such as redshift, black hole mass, etc., they unfortunately smooth out any possible variation between the three images due to the gravitational time delay. Studies of the time-variability of multiply-imaged AGN is important, for example allowing for cosmography studies and reverberation-mapping. This requires careful calibration of the individual spectra though and goes beyond the scope of this work.

\noindent The final stacked spectrum of A2744-QSO1 displays numerous emission lines and is shown in Fig.~2. For the most robust emission line in our spectrum, the Balmer line H$\beta$, we use \texttt{specutils} \cite{specutils23} (\texttt{v1.10.0}) to fit a two-component (broad and narrow) Gaussian profile to the continuum-subtracted emission line and estimate the uncertainties with an MCMC-analysis using \texttt{emcee} \cite{foreman-mackey13} (\texttt{v3.1.4}). Our line-fit explicitly includes the NIRSpec point-source line-spread-function (LSF) \cite{degraaff23} into the model at the wavelength of H$\beta$ by convolving the model line profile with the LSF in each step of the MCMC before calculating the likelihood. The $\mathrm{FWHM}=2800\pm250\,\frac{\mathrm{km}}{\mathrm{s}}$ used in the following is therefore intrinsic, i.e. corrected for instrumental effects. H$\alpha$ is more challenging to model as the red wing of the spectrum falls off the detector and is blended with [N\,\textsc{ii}]\,$\lambda\lambda6550,6585$\AA. In addition, the spectral resolution appears to be high enough at these wavelengths to resolve the narrow component on top of the broad wing. For completeness, we therefore proceed to model H$\alpha$, [N\,\textsc{ii}]\,$\lambda\lambda6550$, H$\beta$, and [O\,\textsc{iii}] jointly, fixing the metal lines to $100\,\frac{\mathrm{km}}{\mathrm{s}}$ FWHM, restricting the narrow-to-broad line ratio to be the same for H$\beta$ and H$\alpha$, but allowing for different line widths between H$\beta$, and H$\alpha$. Encouragingly, the fit parameters are well constrained and reproduce the broad line width of H$\beta$, while finding a smaller FWHM for the broad component of H$\alpha$ $\mathrm{FWHM}\simeq2300\pm250\,\frac{\mathrm{km}}{\mathrm{s}}$ as is commonly seen in low-redshift AGN \cite{greene05}. For consistency, we estimate the width of the narrow lines by fitting the [O\,\textsc{iii}]\,$\lambda5008$\AA\ line with the same method as above and obtain an LSF-corrected line width that is well below the width of the LSF. This shows that the forbidden lines are indeed narrow. Other spectral lines are identified and modeled with single Gaussians and presented in Tab.~\ref{tab:emission lines}. Exceptions to this are H$\gamma$, which is blended with [O\,\textsc{iii}]$\lambda4363$\AA, and Si\,\textsc{ii}\,$\lambda6347$\AA, which is blended with [O\,\textsc{i}]$\lambda6364$\AA\ and [Fe\,\textsc{x}]$\lambda6375$\AA. These are modeled similarly to H$\alpha$, i.e. including the blended emission lines, the LSF and the fixed narrow line width. Note that H$\gamma$ also has a resulting a broad FWHM, consistent with H$\alpha$ and H$\beta$.

We derive a Balmer decrement of $7.4\pm0.4$ from the Balmer line measurements in Tab.~\ref{tab:emission lines}. Assuming an reddening law based on fits to the Small Magellanic Cloud \cite{gordon03} (SMC) and for $R_V=2.72$ \cite{prevot84,bouchet85} we find $A_V=3.0\pm0.5$, $A(\mathrm{H}\beta$)$=3.1\pm0.5$, and $A(\mathrm{H}\alpha)=2.1\pm0.5$. To date, there have been no direct measurements of the dust attenuation law in obscured high-redshift AGN. The SMC law has however been found to match low-redshift dusty AGN best \cite{richards03,hopkins04} and also well matches low-metallicity high-redshift galaxies \cite{capak15,reddy15,reddy18a,shivaei20}. We note that if we assumed a flatter reddening law, the reddening corrections would be larger as would the bolometric luminosity \cite{salim20}. For example, assuming a Calzetti attenuation law \cite{calzetti00}, we obtain $A_V=4.5$. The line fluxes are corrected for dust attenuation before black hole mass and Eddington ratio calculations are made. While the Balmer decrement in the broad-line region arises in part within the broad-line region itself \cite{korista04}, we note that we derive very comparable $A_V$ from the photometric modeling performed in previous work \cite{furtak23d,labbe23}.

To investigate a possible SFR origin of the Balmer lines, we convert the dust-corrected H$\alpha$ luminosity to an SFR \cite{kennicutt98}, resulting in an on-going SFR of $\psi\sim40\,\frac{\mathrm{M}_{\odot}}{\mathrm{yr}}$ and predict the rest-frame far-infrared flux using the \texttt{Flexible Stellar Population Synthesis} code \cite{conroy09}, which implements the Draine \& Li \cite{draine01} dust model. The predicted ALMA 1.2\,mm fluxes are $f_{1.2\,\mathrm{mm}}\sim1$\,mJy in contrast with a $3\sigma$ upper limit of $<0.1$\,mJy based on the deep ALMA map, clearly ruling-out the possibility of star-formation as the origin of the emission lines.

\noindent We use the H$\beta$ and H$\alpha$ line widths and attenuation-corrected line fluxes to estimate the single-epoch black hole mass. Because we do not measure the intrinsic continuum, we use the de-reddened and de-magnified Balmer line luminosity to calculate $M_{\rm BH}$ \cite{greene05}. We find a black hole mass of $M_{\rm BH}=4_{-1}^{+2}\times10^7\,\mathrm{M}_{\odot}$, taking the average of H$\alpha$ and H$\beta$, which are consistent with each other. The errors reflect both the line width uncertainties, the differences between the H$\alpha$ and H$\beta$ measurement and the dust attenuation uncertainty. The additional systematic uncertainty term on single-epoch black hole masses is estimated to be $\sim0.5$\,dex \cite{shen13}. We find a bolometric luminosity of $L_{\mathrm{bol}}=(1.1\pm0.3)\times10^{45}\,\frac{\mathrm{erg}}{\mathrm{s}}$, assuming a bolometric correction of $10\pm2$ \cite{richards06}, based on the emission line luminosities. For consistency, we also derive the bolometric luminosity from the continuum at rest-frame 5100\,\AA\ and find a consistent result. The continuum luminosity estimated from the broad-band photometry in \cite{furtak23d} is higher than this estimate by nearly a factor of ten, resulting from uncertainties in quantities like the redshift and the reddening. We thus infer an Eddington ratio of $L_{\rm bol}/L_{\rm Edd}\sim0.3$.

\subsection*{Size measurement}
Furtak et al. \cite{furtak23d} used \texttt{galfit} \cite{peng10} measurements to the individual images of A2744-QSO1, in all available broad bands from F115W to F444W, together with the fact that all three images appear as point sources despite suffering different amounts of shear and magnification, to conclude that the object was indeed point-like and from the PSF size to obtain a source size estimate of $r_e<35$\,pc \cite{furtak23d}.

\noindent Here we revisit this size estimate with the most up-to-date PSF models and data reduction. Using \texttt{pysersic} \cite{pasha23}, we fit a single S\'{e}rsic model to the F150W image. We find a firm upper limit on the size of $r_{e}<11.2^{+1.6}_{-0.8}$\,milli-arcseconds, which translates to a $2\sigma$ upper limit of $r_e\lesssim100$\,pc for the lensed image. Applying a lensing correction of $\mu_{\mathrm{t}}=3.52$ -- the tangential shear in image~A used for the size measurement --  we then confirm a firm $2\sigma$ upper limit of 30\,pc on the source radius, in agreement with the Furtak et al. measurement \cite{furtak23d}.

\subsection*{Number density}
In order to derive the number density of objects like A2744-QSO1, it is needed to compute the corresponding effective co-moving volume $V_{\mathrm{eff}}$ which folds-in the lensing distortion and possible selection effects. This is done following a forward-modeling approach \cite{atek18}, which was re-implemented for JWST observations and in particular for UNCOVER \cite{chemerynska23}: We use the (de-magnified) SED of A2744-QSO1 to populate the UNCOVER source plane of A2744 \cite{furtak23c} with 10000 similar mock sources. These are then deflected into the lens plane and injected in the UNCOVER mosaics on which we re-run the source detection and selection methods used for the catalogs \cite{weaver23,atek23b}. The ratio of recovered to injected sources allows us to determine the selection-function for objects such as A2744-QSO1. This selection-function is multiplied with the co-moving volume element which is then integrated over the source-plane area of sufficient magnification that allows us to detect the source at $3\sigma$ in each NIRCam band, given the UNCOVER mosaic depths \cite{weaver23}. The thus obtained effective volume for A2744-QSO1 is $V_{\mathrm{eff}}=27541\pm294\,\mathrm{Mpc}^3$.

\noindent Given that the $\sim45\,\mathrm{arcmin}^2$ of the UNCOVER field-of-view \cite{bezanson22} contains two ``red dot'' objects of the redshift and UV luminosity of A2744-QSO1 \cite{labbe23}, A2744-QSO1 itself and one other \cite{greene23} projected $369$\,kpc away in the source plane, the number density is $\phi=(7.3\pm1.7)\times10^{-5}\,\mathrm{Mpc}^{-3}\,\mathrm{mag}^{-1}$.

\subsection*{Theoretical expectation}
We compare our results to the \texttt{Dark matter and the Emergence of gaLaxies in the ePocH of reIonization} \cite{dayal19,piana21} (\texttt{DELPHI}) semi-analytic model. This is suited to jointly track the assembly of the dark matter, baryonic and black hole components of high-redshift ($z\gtrsim7$) galaxies up to $z\sim40$. Starting at $z\sim7$, we build analytic (binary) merger trees for 700 galaxies equally spaced in log-space in the halo mass range $M_{\mathrm{h}}=10^8-10^{15}\,\mathrm{M}_{\odot}$ using a constant time-step of 30\,Myr. Each of these halos is assigned a number density according to the Sheth-Tormen halo mass function \cite{sheth02} at $z\sim7$ and the number density propagated throughout the merger tree. 

\noindent In this work, we only consider seeding by stellar black hole seeds formed from metal-free (Pop.~III) stars. The first halos (starting leaves) of any halo are seeded with a stellar black hole (of 150\,$\mathrm{M}_\odot$) at $z>13$ assuming that halos collapsing from high ($\gtrsim3.5$)-$\sigma$ fluctuations in the primordial density field are most likely to host such seeds. At any time-step, the initial gas mass is the sum of that brought-in by mergers and smooth accretion. A fraction of this gas can form stars with an efficiency that is the minimum between the Supernovae (SN~II) energy required to unbind the rest of the gas and an upper threshold $f_*$. In terms of black holes, in our {\it fiducial} model, these grow both through (instantaneous) mergers and accretion. A fraction of the gas mass left in the halo after star-formation and its associated feedback can be accreted onto the black hole with the accretion rate depending on a critical halo mass (which at $z\sim7$ corresponds to $10^{10.9}\,\mathrm{M}_{\odot}$). While black holes in lower mass halos are assumed to have puffed-up disks, only allowing for low accretion rates ($\sim7.5\times10^{-5}$~Eddington), black holes in halos above this critical mass can accrete a fixed fraction of the gas mass up to the Eddington limit; a fraction of the black hole energy (0.3\,\%) is allowed to couple to the gas content. The {\it maximal (no feedback)} model presents an upper limit where every black hole is always allowed to accrete the gas mass available, up to the Eddington limit and we ignore gas mass lost in outflows, both due to type~II supernova and black hole feedback.

\noindent The model has been tuned to reproduce all the key observables for both star-forming galaxies at $z\sim5-10$ (including the evolving UV luminosity and stellar mass functions) and AGN at $z\sim5-6$ (including the AGN UV luminosity and black hole mass functions).

% ------------------ Backmatter -----------------------------------------
\backmatter

%\bmhead{Supplementary information}

\bmhead{Acknowledgments} We warmly thank the anonymous referees for their comments and suggestions which greatly helped improving this paper. A.Z. and L.J.F. acknowledge support by grant 2020750 from the United States-Israel Binational Science Foundation (BSF) and grant 2109066 from the United States National Science Foundation (NSF), and by the Ministry of Science \& Technology, Israel. H.A. and I.C. acknowledge support from CNES, focused on the JWST mission, and the Programme National Cosmology and Galaxies (PNCG) of CNRS/INSU with INP and IN2P3, co-funded by CEA and CNES. P.D. acknowledges support from the NWO grant 016.VIDI.189.162 (``ODIN'') and the European Commission's and University of Groningen's CO-FUND Rosalind Franklin program. S.F. acknowledges the support from NASA through the NASA Hubble Fellowship grant HST-HF2-51505.001-A, awarded by the \textit{Space Telescope Science Institute} (STScI). The Cosmic Dawn Center (DAWN) is funded by the Danish National Research Foundation under grant No.\ 140. Support for the program JWST-GO-2561 was provided through a grant from the STScI under NASA contract NAS 5-03127. This work has received funding from the Swiss State Secretariat for Education, Research and Innovation (SERI) under contract number MB22.00072, as well as from the Swiss National Science Foundation (SNSF) through project grant 200020\_207349.
     
This work is based on observations obtained with the NASA/ESA/CSA JWST, retrieved from the \texttt{MAST} at the STScI, which is operated by the Association of Universities for Research in Astronomy, Inc. under NASA contract NAS 5-26555. This work is also based on observations made with ESO Telescopes at the La Silla Paranal Observatory and the \textit{Atacama Large Millimeter/sub-millimeter Array} (ALMA), obtained from the ESO Science Archive facility. ALMA is a partnership of ESO (representing its member states), NSF (USA) and NINS (Japan), together with NRC (Canada), MOST and ASIAA (Taiwan), and KASI (Republic of Korea), in cooperation with the Republic of Chile. The Joint ALMA Observatory is operated by ESO, AUI/NRAO and NAOJ. The National Radio Astronomy Observatory (NRAO) is a facility of the NSF operated under cooperative agreement by Associated Universities Inc.

This work was made possible by utilizing the \texttt{CANDIDE} cluster at the Institut d’Astrophysique de Paris, which was funded through grants from the PNCG, CNES, DIM-ACAV, and the Cosmic Dawn Center and is maintained by S.~Rouberol. Cloud-based data processing and file storage for this work is provided by the AWS Cloud Credits for Research program.

\bmhead{Author contributions} A.Z., J.E.G. and L.J.F. wrote the manuscript. L.J.F. made the figures. L.J.F. and I.L. performed the line fits. I.L. designed the program and produced the deep spectrum. R.B. and I.L reduced the data. L.J.F. and A.Z. constructed the lens model. P.D. provided simulations to contextualise the observational results obtained. V.K. and I.L. measured emission line strengths. I.C. ran the completeness simulations. T.B.M., D.S. and E.N. measured the source size. I.L and R.B. are the PIs of the UNCOVER program. All authors contributed to the manuscript and aided the analysis and interpretation.

\bmhead{Data availability} The raw JWST data used in this work are publicly available on the \texttt{Barbara A. Mikulski Archive for Space Telescopes} (\texttt{MAST}). These observations specifically are associated with the JWST GO program number 2561, JWST ERS program number 1324, and JWST DD program number 2756. The reduced UNCOVER mosaics, catalogs, lens models as well as spectra are part of the public data release by the UNCOVER team \cite{bezanson22,furtak23c,weaver23} (see also Price et al. in prep.). The public ALMA data can be found on the ESO Science Archive under IDs 2018.1.00035.L and 2013.1.00999.S. The deep \textit{Chandra} image used here comprises data from over 60 individual programs which are available on the \textit{Chandra Data Archive}. The individual observations IDs are listed in Table~1 in Bogd\'{a}n et al. \cite{bogdan23}.

\bmhead{Code availability} This research made use of \texttt{Astropy}, a community-developed core Python package for Astronomy \cite{astropy13,astropy18} as well as the packages \texttt{NumPy} \cite{vanderwalt11}, \texttt{SciPy} \cite{virtanen20}, \texttt{Matplotlib} \cite{hunter07} and the \texttt{MAAT} Astronomy and Astrophysics tools for \texttt{MATLAB} \cite{maat14}.

\bmhead{Competing interests} The authors declare no competing financial interests.

% Bibliography

%\bibliography{references}

\bmhead{Additional Information} Correspondence and requests regarding reprints and additional information should be addressed to \url{furtak@post.bgu.ac.il}.

\end{document}